\documentclass[aps,pre,preprint,showkeys]{revtex4-1}

\usepackage{graphicx}
\usepackage{dcolumn}
\usepackage{bm}
\usepackage{amsmath}

\def\grl{Geophysics Research Letters}

\begin{document}


\title{Non-steady state model of global temperature change:
Can we keep temperature from rising more than on two degrees?}


\author{Alexei V. Karnaukhov*}
\author{Elena V. Karnaukhova}
\affiliation{Institute of Cell Biophysics, Russian Academy of
Sciences, Puschino, Russian Federation}

\author{Elena P. Popova}
\affiliation{Astronomy Research Center, Bernardo O'Higgins University, Santiago, Chile}

\author{Mikhail S. Blinnikov}
\affiliation{Geography and Planning Department, St. Cloud State University, St. Cloud, MN United States}

\author{Konstantin A. Shestibratov}
\affiliation{M. A. Shemyakin and Yu. A. Ovchinnikov Institute of
Bioorganic Chemistry, \\
Russian Academy of Sciences, Moscow, Russian Federation}

\author{Sergei I. Blinnikov}
\affiliation{Institute for Theoretical and Experimental Physics, National Research Center "Kurchatov Institute", Moscow, Russian Federation}

\author{Vladimir N. Reshetov}
\affiliation{Institute for Laser and Plasma Technologies, Moscow Engineering Physics Institute, Moscow, Russian Federation}

\author{Sergei F. Lyuksyutov*}
\affiliation{Physics Department, University of Akron, Akron OH, United States}

\email{sfl@uakron.edu; alexeikarnaukhov@yandex.ru}

\date{\today}

\begin{abstract}

We propose a non-steady state model of the global temperature change. The model describes Earth's surface temperature dynamics under main climate forcing. The equations were derived from basic physical relationships and detailed assessment of the numeric parameters used in the model. It shows an accurate fit with observed changes in the surface mean annual temperature (MAT) for the past 116 years. Using our model, we analyze the future global temperature change under scenarios of drastic reductions of CO\textsubscript{2}. The presence of non-linear feed-backs in the model indicates on the possibility of exceeding two degrees threshold even under the carbon dioxide drastic reduction scenario. We discuss the risks associated with such warming and evaluate possible benefits of developing CO\textsubscript{2}-absorbing deciduous tree plantations in the boreal zone of Northern Hemisphere.

\end{abstract}

\pacs{}
\keywords{Climate sensitivity; non-steady state climate model; global temperature; albedo}


\maketitle

\section{\label{intro} Inroduction}

The need to plan specific mitigation measures for the expected global rise in land surface
temperatures requires basic understanding of the climate change science for its presentation in a
simple format, accessible to decision making institutions and governments worldwide. The
standard approach is to use the results of 3D general circulation models (GCMs) using an
ensemble of expected temperature changes under, for example, doubling of CO2, or expected
change in the concentrations of all major greenhouse gases (GG). The IPCC report \cite{IPCCWG1PhysicalStocker2013} has not
changed the expected global Earth surface temperature rise of
$\Delta T_{\rm 2xCO_2} = (1.5 \div 4.5)$\,K.

This prediction can be written in a logarithmic form as:
\begin{equation}
 \Delta T(t) = \sigma_{\rm IPCC} \cdot \ln\left( \frac{\rho_{\rm CO_2}(t)}{\rho_{\rm CO_2}(t_0)}  \right)
\label{DeltaT}
\end{equation}
where $\Delta T(t)$ -- average global Earth surface temperature relative to the
pre-industrial levels, $\Delta T(t)\equiv T(t)-T(t_0)$;

\noindent $\rho_{\rm CO_2}(t_0)$
{}-- pre-industrial CO\textsubscript{2 }concentration;

\noindent $\rho_{\rm CO_2}(t)$
{}-- observed level of CO\textsubscript{2} concentration at a given moment of time \textit{t};

\noindent $\sigma_{\rm IPCC} \equiv \frac{\Delta T_{\rm 2xCO_2}}{\ln 2} =  (2.16 \div 6.5)$\,K
{}-- climate sensitivity parameter estimated from models.

A ``canonical''
form of the equation (\ref{DeltaT}) does not take into account some important
facts known about the Earth climate system. A major weakness of GCMs is
their inability to adequately quantify certain feed-backs, although the
models are constantly improving. For example, Meraner et al.~\cite{Meraner2013}
note that the latest generation of climate models consistently
exhibits an increasing equilibrium climate sensitivity (ECS) in warmer
climates due to a strengthening of the water-vapor feedback with
increasing surface temperatures. The increasing ECS is replicated in
their work as a one-dimensional radiative-convective equilibrium model,
which further shows that the enhanced water-vapor feedback follows from
the rising of the tropopause in a warming climate. This feedback challenges the
notion of linear radiative response of the Earth climate.
We propose a non-steady state model. Colman and McAvaney~\cite{Colman2009} found that, as climate
warms, climate sensitivity weakens (although not monotonically); albedo
feedback weakens; water vapor feedback strengthens; and lapse rate
feedback increases negatively. The latter change essentially
offsets the water vapor increase. Ingram~\cite{Ingram2013} proposed a
new approach to quantifying the non-cloud long wave (LW) water vapor
feed-backs that avoids the common problem of the feedback breakdown into
lapse rate change and direct water vapor radiative feedback near
ground. We follow these authors to design a simple, yet
robust, model of the Earth temperature change that accounts
CO\textsubscript{2}, thermal inertia of climate system, water vapor,
and albedo effects. We also use our earlier results
\cite{Karnaukhov2001,Karnaukhov2008}, in which we found precise
solutions to the differential greenhouse effect in the optically dense
atmosphere if the surface input is negligibly small.

Our approach differs from many other in the usage of a renormalization
calculation of climate model parameters that allows to explicitly
include water vapor feedback \cite{Karnaukhov2006}
that is  ``typically neglected'' \cite{Feldl2013}. 
Not only this increases the value of
MAT, but does also lengthen the relaxation time required for the temperature curve leveling off.
Our results show the substantially
non-steady state response of the Earth's climate to the
current forcing that can be clearly observed.

In this work, we consider a simple zero-dimensional non-steady state model (see Eq.(\ref{model}) below
in Sec.\ref{sec:paramModel})
that can better describe the general sensitivity of the Earth's
temperature to the change of CO\textsubscript{2} concentrations using
global feedbacks  \cite{Feldl2013}. 
Unlike the spatially
explicit GCMs, this conceptual model's strength is its
applicability as a working tool to a wide range of climate policy
decision making institutions and climate change mitigation planning.

\section{\label{sec:results} Model and its Results}

\subsection{\label{sec:paramModel} Main parameters of the model}

In terms of major parameters, our model can be simply introduced as the
following differential equation:

\begin{equation}
\frac{d}{dt}  \Delta T(t) = \frac{1}{\tau} \cdot \left( \sigma_{\rm CO_2} \cdot
\ln\left( \frac{\rho_{\rm CO_2}(t)}{\rho_{\rm CO_2}(t_0)}  \right) +
\sigma_a \cdot  \left( \frac{\Delta S(t)}{S} - \frac{\Delta a(t)}{1-a} \right)
- \Delta T(t)  \right) ,
\label{model}
\end{equation}
where new parameters relative to Eq.~(\ref{DeltaT}) describe:

\noindent
$ \Delta a(t) = a(t) - a_0(t) $ -- change in the spherical albedo of the Earth
($ a \simeq  a(t_0) \simeq a(t)$, $\Delta a(t) \ll a $);

\noindent
$ \Delta S(t) = S(t) - S_0(t) $ -- change in the solar constant (power of Solar radiation falling on
top of the Earth's atmosphere $ S \simeq  S(t_0) \simeq S(t)$, \hspace{2mm} $\Delta S(t) \ll S $);

\noindent $\sigma_{\rm CO_2}$  -- climate sensitivity to the change in
CO\textsubscript{2 }concentration;

\noindent $\sigma_a$  -- climate sensitivity to the change in the albedo;

\noindent $\tau$ -- relaxation time constant of the Earth’s climate system.

This model is described by only three key global climate parameters: $\sigma_{\rm CO_2}$, $\sigma_a$, and
$\tau$, which are generally considered most significant in determining
temperature change \cite{Atwood2016,IPCCWG1PhysicalStocker2013}.

The starting point of the main equations of the non-stationary model of
global climate is the condition of overall conservation of energy
within the Earth system:
\begin{equation}
 \frac{d T(t)}{dt}  = \frac{1}{C}  \cdot \left( W_{\rm in}^S (t) - W_{\rm out}^T (t) \right),
\label{Tequation}
\end{equation}
where $C$ is surface heat capacity,
$W_{\rm in}^S (t)$ -- power of incoming short-wave solar radiation,
$W_{\rm out}^T (t)$ -- power of outgoing infrared radiation.

The equation (\ref{Tequation}) can be used both for global estimate of the total heat
capacity and power, and also for deriving average values per unit of
area. The most common application is the latter approach. In this case,
the power of incoming radiation $W_{\rm in}^S (t)$
 and outgoing radiation $W_{\rm out}^T (t)$
can be represented as:
\begin{subequations}
\begin{equation}
W_{\rm in}^S (t)= (1 - a(t))S(t) ;
\label{Win}
\end{equation}
\begin{equation}
W_{\rm out}^T (t) = \sigma T_{\rm eff}^4(t) ,
\label{Wout}
\end{equation}
\end{subequations}
where $S(t)$ -- solar constant;
$a(t)$ -- spherical albedo;
$\sigma$ -- Boltzmann constant;
$T_{\rm eff}(t)$ -- effective Earth temperature.
The latter $T_{\rm eff}(t)$
differs from the average global surface temperature $T(t)$
by the amount of greenhouse effect $\Delta T_G(t)$
by definition:
\begin{equation}
T_{\rm eff}(t) = T(t) - \Delta T_G(t) .
\label{defTG}
\end{equation}

The amount of greenhouse effect can in turn be expressed through the
concentration of greenhouse gases (GG), especially CO\textsubscript{2} and water vapor
H\textsubscript{2}O:
\begin{equation}
 \Delta T_G(t) =  \Delta T_G(t_0) +
 \sigma'_{\rm CO_2} \cdot \ln\left( \frac{\rho_{\rm CO_2}(t)}{\rho_{\rm CO_2}(t_0)}  \right) +
 \sigma'_{\rm H_2O} \cdot \ln\left( \frac{\rho_{\rm H_2O}(t)}{\rho_{\rm H_2O}(t_0)}  \right) + \ldots ,
\label{dotTG}
\end{equation}
where the corresponding values at the moment of time $t_0$
are considered as pre-industrial $\Delta T_G(t_0) \approx 35^\circ$C, $\rho_{\rm CO_2}(t_0)\approx280$~ppm
\cite{Mann1998}. The parametrization of the greenhouse
gases effect as in (\ref{dotTG}) not only fits the traditional approach of (\ref{DeltaT}),
but also is a result of theoretical assessment given by both the
radiation-convection models for small values of the differential
greenhouse effect, $ |\Delta T_G(t) - \Delta T_G(t_0) | < \Delta T_G(t_0) $,
and also the radiation-adiabatic model, describing a strong
greenhouse effect in the optically dense atmospheres
\cite{Blinnikov1996,Blinnikov1997,Karnaukhov2001,Karnaukhov2008}. We
will also note that the parameter $ \sigma'_{\rm CO_2}$
in (\ref{dotTG}) is a basic value of the temperature sensitivity climate
constant, which can be very different from the renormalized $ \sigma_{\rm CO_2}$,
used in the equation (\ref{model}).

\subsection{\label{sec:Derivation} Derivation of the non-steady state
model main equations of global climate considering an impact of the positive water vapor
feedback}

In order to move from the condition of energy balance to the model of
(\ref{model}) we must, first of all, set the starting moment of time $t_0$
(pre-industrial period), when the incoming and outgoing radiation were
balanced, $ W_{\rm in}^S (t_0) - W_{\rm out}^T (t_0)$,
and we can then formulate the energy balance equation (\ref{Tequation})  as:
\begin{equation}
 \frac{d T(t)}{dt}  = \frac{1}{C}  \cdot \left( ( W_{\rm in}^S (t)-W_{\rm in}^S (t_0)) -
 (W_{\rm out}^T (t) - W_{\rm out}^T (t_0)) \right),
\label{Tequation7}
\end{equation}

Using expressions (\ref{Win}) and (\ref{Wout}) for $W_{\rm in}^S (t_0)$ and $W_{\rm out}^T (t_0)$
 and considering that the changes of these corresponding parameters are
small, we may present (\ref{Tequation7}) as:
\begin{eqnarray}
 \frac{d T(t)}{dt}  = \frac{1}{C}  \cdot \left( ( (1 - a(t))S(t)-(1 - a(t))S(t_0)) -
 ( \sigma T_{\rm eff}^4(t) - \sigma T_{\rm eff}^4(t_0)) \right) \nonumber \\
 = \frac{1}{C}  \cdot \left( ( (1 - a)S)\cdot\left(\frac{\Delta S(t)}{S} - \frac{\Delta a(t)}{1-a} \right) -
 4\sigma T_{\rm eff}^3\cdot \Delta T_{\rm eff}(t) \right),
\label{Tequation8}
\end{eqnarray}
where we introduce new values of $T_{\rm eff}$
and $\Delta T_{\rm eff}(t)$: $T_{\rm eff} \simeq T_{\rm eff}(t_0) \simeq T_{\rm eff}(t)$,
$\Delta T_{\rm eff}(t) = T_{\rm eff}(t)-T_{\rm eff}(t_0) \ll T_{\rm eff} $,
analogous to the earlier ones for albedo and the solar constant used in (\ref{model}).

Using (\ref{Win}) and (\ref{Wout}) and introducing
$ W_0 \simeq W_{\rm in}^S (t_0) \simeq W_{\rm out}^T (t_0)$,
equation (\ref{Tequation8}) can be expressed as:
\begin{equation}
 \frac{d T(t)}{dt}
 = \frac{1}{C}  \cdot \left(  W_0\cdot\left(\frac{\Delta S(t)}{S} - \frac{\Delta a(t)}{1-a} \right) -
 4 \frac{W_0}{T_{\rm eff}} \cdot \Delta T_{\rm eff}(t) \right).
\label{Tequation9}
\end{equation}
It should be noted that $\Delta T_{\rm eff}(t)$
in (\ref{Tequation8}) and having (\ref{defTG}) in mind may be expressed through the magnitude
of greenhouse effect $\Delta T_G(t)$, which in turn depends on the concentration of the main greenhouse
gases (\ref{dotTG}) (in this paper we only consider the two main greenhouse gases --
CO\textsubscript{2} and water vapor):
\begin{eqnarray}
 \Delta T_{\rm eff}(t) = T_{\rm eff}(t) - T_{\rm eff}(t_0)  =
 \left(T(t) - \Delta T_G(t)\right) - \left(T(t_0) - \Delta T_G(t_0) \right)  = \\ \nonumber
 = T(t) - T(t_0) - \sigma'_{\rm CO_2} \cdot \ln\left( \frac{\rho_{\rm CO_2}(t)}{\rho_{\rm CO_2}(t_0)}  \right)
 - \sigma'_{\rm H_2O} \cdot \ln\left( \frac{\rho_{\rm H_2O}(t)}{\rho_{\rm H_2O}(t_0)}  \right) .
\label{dTeff}
\end{eqnarray}

With the help of this equation for $ \Delta T_{\rm eff}(t)$, and placing the multiplier $4W_0/T_{\rm eff}$
outside the parentheses and keeping in mind that
$dT(t)/dt = d \Delta T(t)/dt$, the equation (\ref{Tequation9}) can be presented as:
\begin{equation}
 \frac{d \Delta T(t)}{dt}
 = \frac{4W_0}{CT_{\rm eff}} \left( \frac{T_{\rm eff}}{4}
 \left(\frac{\Delta S(t)}{S} - \frac{\Delta a(t)}{1-a} \right) -
 \Delta T(t) + \sigma'_{\rm CO_2} \ln\left( \frac{\rho_{\rm CO_2}(t)}{\rho_{\rm CO_2}(t_0)}  \right)
 + \sigma'_{\rm H_2O} \ln\left( \frac{\rho_{\rm H_2O}(t)}{\rho_{\rm H_2O}(t_0)}  \right) \right) .
\label{dotDeltaT}
\end{equation}
Using now:
\begin{subequations}
\begin{equation}
 \tau' \equiv \frac{CT_{\rm eff}}{4W_0} ;
 \label{tauPrim}
\end{equation}
\begin{equation}
 \sigma_a' \equiv \frac{T_{\rm eff}}{4} ,
 \label{sigmaPrim}
\end{equation}
\end{subequations}
and changing the order in the right-hand side of the equation (\ref{dotDeltaT}), we obtain
the resulting equation for $\Delta T(t)$
that differs from the basic equation (\ref{model}) only by having the parameter
that expresses the water vapor feedback:
\begin{equation}
 \frac{d}{dt} \Delta T(t)
 = \frac{1}{\tau'} \left(  \sigma'_{\rm CO_2} \ln\left( \frac{\rho_{\rm CO_2}(t)}{\rho_{\rm CO_2}(t_0)}  \right)
 + \sigma'_{\rm H_2O} \ln\left( \frac{\rho_{\rm H_2O}(t)}{\rho_{\rm H_2O}(t_0)}  \right)
 + \sigma'_a \cdot  \left(\frac{\Delta S(t)}{S} - \frac{\Delta a(t)}{1-a} \right) -
 \Delta T(t)  \right) .
 \label{dotDelTH2O}
\end{equation}
We note that unlike other dynamic parameters in (\ref{dotDelTH2O}), water vapor
 concentration in the atmosphere $\rho_{\rm H_2O}$
 strongly depends on the mean air temperature (MAT) and can be excluded
 from (\ref{dotDelTH2O}). In order to do that, we must first express saturated water
 vapor pressure $\rho^S_{\rm H_2O}$, as a function of temperature, using the Boltzmann
 distribution:
 \begin{equation}
  \rho^S_{\rm H_2O}(T)=\rho^L_{\rm H_2O} \cdot \exp\left( - \frac{\Delta E}{R\cdot T} \right),
  \label{rhoS}
 \end{equation}
 where $\rho^L_{\rm H_2O}$ is the density of liquid water, $\Delta E$
 -- molar energy of the phase change liquid to gas, 
 $R$ -- universal gas constant.
 Second, we need to use a linear
 approximation to describe the dependence of the mean relative humidity $\delta \left( T(t) \right) $
 on the change in MAT:
 \begin{equation}
  \delta \left( T(t) \right) = \frac{ \rho_{\rm H_2O}(t)}{  \rho^S_{\rm H_2O}(T(t))}
  \simeq \delta_0 - \delta_1 \cdot \Delta T .
  \label{delTt}
 \end{equation}
Indeed, using (\ref{delTt}) the parameter that is dependent on $\rho_{\rm H_2O}(t) $
is expressed as:
\begin{eqnarray}
 \sigma'_{\rm H_2O} \cdot \ln\left( \frac{\rho_{\rm H_2O}(t)}{\rho_{\rm H_2O}(t_0)}  \right) = \nonumber
 \sigma'_{\rm H_2O} \cdot \ln\left( \frac{(\delta_0 - \delta_1\cdot\Delta T(t)) \cdot
 \rho^S_{\rm H_2O}(T(t_0)+\Delta T(t))}{\delta_0 \rho^S_{\rm H_2O}(T(t_0))}  \right) =  \\ \nonumber
= \sigma'_{\rm H_2O} \cdot \ln\left( \left( \frac{\delta_0 - \delta_1\cdot\Delta T(t)}{\delta_0} \right)
\cdot
 \frac{ \rho^S_{\rm H_2O}(T(t_0)+\Delta T(t))}{ \rho^S_{\rm H_2O}(T(t_0))}  \right) = \\
= \sigma'_{\rm H_2O} \cdot \left( \ln\left( \frac{\delta_0 - \delta_1\cdot\Delta T(t)}{\delta_0} \right)
+ \ln \left(
 \frac{ \rho^S_{\rm H_2O}(T(t_0)+\Delta T(t))}{ \rho^S_{\rm H_2O}(T(t_0))}  \right) \right) .
\label{sigmaH2O}
\end{eqnarray}
We can now transform (\ref{sigmaH2O}) by using linear approximation: $ \ln(1+k\cdot x \simeq k \cdot x $,
and substitution of (\ref{rhoS}) for $\rho^S_{\rm H_2O}(T)$  into the equation (\ref{sigmaH2O}):
\begin{eqnarray}
 \sigma'_{\rm H_2O} \cdot \ln\left( \frac{\rho_{\rm H_2O}(t)}{\rho_{\rm H_2O}(t_0)}  \right) \simeq \nonumber
 \sigma'_{\rm H_2O} \cdot \left( \ln\left( \frac{ \rho^L_{\rm H_2O} \cdot
 \exp\left(- \frac{\Delta E}{R\cdot(T(t_0)+\Delta T(t))} \right)
 }{ \rho^L_{\rm H_2O} \cdot
 \exp\left(- \frac{\Delta E}{R\cdot(T(t_0))} \right)} \right)
 - \frac{\delta_1}{\delta_0}\cdot\Delta T(t)  \right) =  \\
=  \sigma'_{\rm H_2O} \cdot \left( \ln\left(
 \exp\left(\frac{\Delta E}{R\cdot T(t_0)} - \frac{\Delta E}{R\cdot(T(t_0)+\Delta T(t))} \right) \right)
 - \frac{\delta_1}{\delta_0}\cdot\Delta T(t)  \right) .
\label{sigmapH2O}
\end{eqnarray}

With $\ln(\exp(x))=x$ and $\frac{1}{1+x} \simeq 1 -x $, we finally arrive at the relatively simple expression:
\begin{eqnarray}
 \sigma'_{\rm H_2O} \cdot \ln\left( \frac{\rho_{\rm H_2O}(t)}{\rho_{\rm H_2O}(t_0)}  \right) \simeq \nonumber
 \sigma'_{\rm H_2O} \cdot \left(
 \frac{\Delta E}{R\cdot T(t_0)} \left(1 - \left(1 - \frac{\Delta T(t)}{T(t_0)} \right) \right)
 - \frac{\delta_1}{\delta_0}\cdot\Delta T(t)  \right) \\
 = \sigma'_{\rm H_2O} \cdot \left(
 \frac{\Delta E}{R\cdot T^2(t_0)}
 - \frac{\delta_1}{\delta_0}\right) \cdot \Delta T(t)  .
\label{sigmaTH2O}
\end{eqnarray}
Using (\ref{sigmaTH2O}), our model of (\ref{dotDelTH2O}), becomes:
\begin{eqnarray}
 \frac{d}{dt} \Delta T(t)
 = \frac{1}{\tau'} \left(  \sigma'_{\rm CO_2} \ln\left( \frac{\rho_{\rm CO_2}(t)}{\rho_{\rm CO_2}(t_0)}  \right) \nonumber
 + \sigma'_{\rm H_2O} \left(
 \frac{\Delta E}{R\cdot T^2(t_0)}
 - \frac{\delta_1}{\delta_0}\right) \Delta T(t) \right. \\
 \left. + \sigma'_a   \left(\frac{\Delta S(t)}{S} - \frac{\Delta a(t)}{1-a} \right) -
 \Delta T(t)  \right) .
 \label{dotDelTH2Op}
\end{eqnarray}
Introducing further:
\begin{equation}
  k \equiv \frac{1}{1-\xi}, \quad \mbox{and} \quad \xi \equiv \sigma'_{\rm H_2O} \left(
 \frac{\Delta E}{R\cdot T^2(t_0)}
 - \frac{\delta_1}{\delta_0}\right) ,
 \label{kxi}
\end{equation}
the equation (\ref{dotDelTH2Op}) may become even simpler:
\begin{eqnarray}
 \frac{d}{dt} \Delta T(t)
 = \frac{1}{\tau'} \left(  \sigma'_{\rm CO_2} \ln\left( \frac{\rho_{\rm CO_2}(t)}{\rho_{\rm CO_2}(t_0)}  \right) \nonumber
  + \sigma'_a   \left(\frac{\Delta S(t)}{S} - \frac{\Delta a(t)}{1-a} \right)
 - (1-\xi) \Delta T(t)  \right) \\
 = \frac{1-\xi}{\tau'} \left( \frac{1}{1-\xi} \cdot \sigma'_{\rm CO_2} \ln\left( \frac{\rho_{\rm CO_2}(t)}{\rho_{\rm CO_2}(t_0)}  \right) \nonumber
  + \frac{1}{1-\xi} \cdot \sigma'_a   \left(\frac{\Delta S(t)}{S} - \frac{\Delta a(t)}{1-a} \right)
 - \Delta T(t)  \right) \\
 = \frac{1}{k\tau'} \left( k \cdot \sigma'_{\rm CO_2} \ln\left( \frac{\rho_{\rm CO_2}(t)}{\rho_{\rm CO_2}(t_0)}  \right)
  + k \cdot \sigma'_a   \left(\frac{\Delta S(t)}{S} - \frac{\Delta a(t)}{1-a} \right)
 - \Delta T(t)  \right) ,
 \label{dotDelTH2Oq}
\end{eqnarray}
which is basically similar to our initial model equation (\ref{model})
considering:
\begin{subequations}
\begin{equation}
 \sigma_{\rm CO_2} \equiv k\sigma'_{\rm CO_2},
 \label{sigmaCO2k}
\end{equation}
\begin{equation}
 \sigma_a \equiv  k\sigma_a',
 \label{sigmaak}
\end{equation}
\begin{equation}
 \tau \equiv k\tau' .
 \label{tauk}
\end{equation}
\end{subequations}
Our model thus expressed highlights the significance of the water vapor
feedback in making climate system highly sensitive to temperature
increases in a non-linear way. Besides enhancing the role of greenhouse
effect of CO\textsubscript{2} and other GG (\ref{sigmaCO2k}),
the feedback also amplifies other climate forcings, such as changes in albedo $\Delta a(t)$
 and the incoming solar insolation $\Delta S(t)$, which results in renormalization
 of the constant of climate sensitivity (\ref{sigmaak}). Therefore, $k$
 plays a role of the amplification coefficient ($\xi$
 -- coefficient of positive feedback).

A major consequence of the model discussed above is the sensitivity of
relaxation time $\tau$, which is basically a measure of the thermal inertia of the climate
system (\ref{tauk}).

\subsection{\label{sec:Estimation} Estimation of numeric values of the key model parameters}

We now have three key parameters in the model (\ref{model}) that must be estimated
numerically: $\sigma_{\rm CO_2}, \sigma_a, \tau$, because they determine MAT dynamic
$\Delta T(t)$, under changing main climate forcings ($\rho_{\rm CO_2}(t), \Delta a(t), \Delta S(t)$).

The main challenge in estimating these values is the estimate of the
coefficient $k$.
This is not the result of complex calculations for $\xi$ (\ref{kxi}),
rather the outcome of the mathematical formula that gives a
major error in $k$ with even slight error in $\xi$
at values $\xi\approx 1$. Moreover, if $\xi \ge 1$
climate system described in the equation (\ref{dotDelTH2Oq}) loses stability and
generates potentially unlimited runaway values for temperature.

The fact of reasonably stable climate system of Earth for the past 4
billion years despite high variability of climate parameters speaks
against such values. In other words, based on the
Earth's track record, neither $\xi$, nor $k$
can be too big, or else the Earth temperature regime as we know it
would have ended long time ago. \ Below we describe a way to estimate
the current values for the three key parameters of the model ($\sigma_{\rm CO_2}, \sigma_a, \tau$)
based on the data from IPCC and NASA.

First of all, we need to estimate the original, non-renormalized climate
sensitivities $\sigma'_{\rm CO_2}, \sigma'_a$,
and time constant $\tau'$.

\subsection{ CO\textsubscript{2}-dependent climate sensitivity constant $\bm{\sigma'_{\rm CO_2}}$ (original value)}

Changing CO\textsubscript{2} concentration
$\rho_{\rm CO_2}(t_0)\rightarrow \rho_{\rm CO_2}(t) $
under constant MAT of the Earth
surface $\Delta T(t)= 0$ , albedo $ \Delta a(t) = 0$ and other factors set as in (\ref{defTG}) and (\ref{dotTG}) will result in a change
in the effective temperature $\Delta T_{\rm eff}$:
\begin{equation}
 \Delta T_{\rm eff}(t) =  - \sigma'_{\rm CO_2} \cdot \ln\left( \frac{\rho_{\rm CO_2}(t)}{\rho_{\rm CO_2}(t_0)}  \right) .
\label{delTeffsig}
\end{equation}

This, in turn, will alter the intensity of the outgoing thermal
radiation $\Delta W_{\rm out}^T$, which according to (\ref{Wout}) and (\ref{delTeffsig}):
\begin{equation}
 \Delta W_{\rm out}^T \simeq \frac{\partial W_{\rm out}^T }{ \partial  T_{\rm eff} }
   \Delta T_{\rm eff} = -4\sigma  T_{\rm eff}^3 \sigma'_{\rm CO_2} \cdot \ln\left( \frac{\rho_{\rm CO_2}(t)}{\rho_{\rm CO_2}(t_0)}  \right) =
   - \frac{4W_{\rm out}^T}{T_{\rm eff}} \cdot \sigma'_{\rm CO_2} \cdot \ln\left( \frac{\rho_{\rm CO_2}(t)}{\rho_{\rm CO_2}(t_0)}  \right) .
\label{delWout}
\end{equation}
From (\ref{delWout}) it follows that the  CO\textsubscript{2}{}-dependent climate
sensitivity constant will be:
\begin{equation}
 \sigma'_{\rm CO_2} = \frac{\Delta W^{\rm RF} \cdot T_{\rm eff}}
    {4W_{\rm out}^T\cdot  \ln\left( \frac{\rho_{\rm CO_2}(t)}{\rho_{\rm CO_2}(t_0)}  \right)  } ,
\label{sigCO2Wout}
\end{equation}
where $ \Delta W^{\rm RF} $ is the radiation forcing. In our case,
$ \Delta W^{\rm RF} = - W_{\rm out}^T $, because the reduction in the intensity of
outgoing thermal radiation,
from the energy balance perspective, is equivalent to an increase in
incoming solar radiation forcing. To estimate $ \sigma'_{\rm CO_2}$ (Table~\ref{tab:table1}) we used the
published values of the parameters in (\ref{sigCO2Wout}) from
IPCC \cite{IPCC1995,IPCCWG1PhysicalStocker2013} ($ T_{\rm eff} \simeq 254$~K is also published by NASA \cite{NASA2016}).

\begin{table}[b]
\caption{\label{tab:table1}%
Key parameters needed to estimate CO\textsubscript{2}-dependent climate
sensitivity.
}
\begin{ruledtabular}
\begin{tabular}{ccccccc}
$t$ & $ \rho_{\rm CO_2}(t_0)$ & $\rho_{\rm CO_2}(t)$ & $ T_{\rm eff}$ & $W_{\rm out}^T$ & $ \Delta W^{\rm RF} $ & $\sigma'_{\rm CO_2}$ \\
(year) & (ppmv)  &  (ppmv)  & (K) &  (W/m$^2$) &  (W/m$^2$) & (K) \\
\colrule
1994 &  280\footnote{IPCC 1995 \cite{IPCC1995}  (pp. 15, 78)} &  358\footnotemark[1]  &  254\footnote{NASA \cite{NASA2016}} \footnote{IPCC 1995 \cite{IPCC1995} (p. 57)} &  235\footnotemark[3] &  1.5\footnote{IPCC 1995 \cite{IPCC1995} (pp. 17, 24, 117)} &   \textbf{1.64}  \\
2011 (min) &
278\footnote{IPCC 2013 \cite{IPCCWG1PhysicalStocker2013} (p. 467)} &
391\footnote{IPCC 2013 \cite{IPCCWG1PhysicalStocker2013} (p. 11)}  &
254\footnotemark[2] &
239\footnote{IPCC 2013 \cite{IPCCWG1PhysicalStocker2013} (p. 181)}  &
1.33\footnote{IPCC 2013 \cite{IPCCWG1PhysicalStocker2013} (p. 14)} &
\textbf{1.03}\\
2011 &
278\footnotemark[5] &
391\footnotemark[6] &
254\footnotemark[2] &
239\footnotemark[7] &
1.68\footnotemark[8] &
\textbf{1.31}\\
2011 (max) &
278\footnotemark[5] &
391\footnotemark[6] &
254\footnotemark[2] &
239\footnotemark[7] &
2.05\footnotemark[8] &
\textbf{1.59}\\
\end{tabular}
\end{ruledtabular}
\end{table}

From Table~\ref{tab:table1}, the estimate of the non-renormalized
CO\textsubscript{2}-dependent climate
sensitivity is:
\begin{equation}
 \sigma'_{\rm CO_2} = (1.31 \pm 0.28)\;\mbox{K}
 \label{climateSens}
\end{equation}

\subsection{\label{sec:albedo} Albedo-dependent climate sensitivity  $\bm{\sigma'_a}$ (non-renormalized value)}

This parameter $\sigma'_a$
 can be directly estimated from (\ref{sigmaPrim}) and the value of $ T_{\rm eff}$
from Table~\ref{tab:table1}:
\begin{equation}
 \sigma'_a =  \frac{T_{\rm eff}}{4} = 63.5\;\mbox{K}.
 \label{albedoSens}
\end{equation}

\subsection{ Relaxation time constant of the Earth climate system $\bm{ \tau'}$
(non-renormalized value)}

This parameter $\tau'$
 is estimated based on (\ref{tauPrim}) and the estimates from
the last IPCC report \cite{IPCCWG1PhysicalStocker2013} (Table~\ref{tab:table2}).
Besides that, we
need to analyze the thermal heat storage capacity of the Earth surface $C$
because in (\ref{tauPrim}) the values are normalized per unit of surface area:
\begin{equation}
 C \simeq   \frac{\Delta Q_{40} }{\Delta T_{40}} \cdot \frac{1}{S_E} =
 \frac{1}{\Delta T_{40}}\cdot \frac{\Delta Q_{40}}{4\pi R_{mE}^2} ,
 \label{CdelQ}
\end{equation}
where $ \Delta Q_{40}$
 - change in thermal heat storage capacity and $\Delta T_{40}$
 - \ change of the average MAT for 40 years (from 1971 to 2010);
 $S_E=4\pi R_{mE}^2 = 5.10\cdot 10^{14}\; \mbox{m}^2${}- surface area of the Earth; \
 $R_{mE} = 6.371\cdot10^6$~m
 - average Earth radius \cite{NASA2016}. 

\begin{table}[b]
\caption{\label{tab:table2}%
Key parameters needed to estimate relaxation time constant.
}
\begin{ruledtabular}
\begin{tabular}{cccccccc}
& $\Delta Q_{40}$ & $ \Delta Q_{40}/A_E$ & $\Delta T_{40}$ & $ C $ & $T_{\rm eff}$ & $ W_0 \simeq W^T_{\rm out} $ & $\tau'$ \\
& (J) & (J/m$^2$)  & (K) &  (J/m$^2$K) & (K)&  (W/m$^2$) & (year) \\
\colrule
min $\tau'$ &  $19.6\cdot 10^{22}$\footnote{IPCC 2013 \cite{IPCCWG1PhysicalStocker2013} (p. 39)} &  $3.84\cdot 10^8 $  &  0.56\footnote{IPCC 2013 \cite{IPCCWG1PhysicalStocker2013} (p. 37)} &  $6.86\cdot 10^8 $ &  254\footnote{NASA \cite{NASA2016}} \footnote{IPCC 1995 \cite{IPCC1995} (p. 57)} & 239\footnote{IPCC 2013 \cite{IPCCWG1PhysicalStocker2013} (p. 181)} &   5.78  \\
$\tau'$ &  $27.4\cdot 10^{22}$\footnotemark[1] &  $5.37\cdot 10^8 $  &  0.48\footnotemark[2] &  $11.2\cdot 10^8 $ &  254\footnotemark[3] \footnotemark[4] & 239\footnotemark[5] &   9.44  \\
max $\tau'$ &  $35.1\cdot 10^{22}$\footnotemark[1] &  $6.88\cdot 10^8 $  &  0.32\footnotemark[2] &  $21.5\cdot 10^8 $ &  254\footnotemark[3] \footnotemark[4] & 239\footnotemark[5] &   18.1  \\
\end{tabular}
\end{ruledtabular}
Note: in geophysical literature it is common
to use years for time constant. Our formula
(\ref{tauPrim}) gives time values in seconds.
Conversion coefficient: $n_\tau =3.16\cdot 10^7$~sec/yr.

\end{table}

Thus, the estimate of the time constant
(relaxation time) $\tau'$
based on the Fifth report of the IPCC \cite{IPCCWG1PhysicalStocker2013} will be:
\begin{equation}
 \tau' = \frac{CT_{\rm eff}}{4W_0}\cdot \frac{1}{n_\tau} \simeq 9.44 \; [5.78 \; \mbox{to}\; 18.1] \;
 \mbox{years} .
 \label{tauPrim2}
\end{equation}

\subsection{Renormalization coefficient \textbf{\textit{k}} }

Direct estimate of coefficient $k$
 (20) is not straightforward. The suggested method includes using
empirical observations of the key climate parameters. For example, an
important fact is the empirical curve of the growth of CO\textsubscript{2} (Fig. 1).

Fig. 1A shows that our model's curve fits the growth of
the observed CO\textsubscript{2} values rather well. The red curve is based on:
 \begin{equation}
  \rho_{\rm CO_2}(t) \simeq \rho_{\rm CO_2}(t_0)
  + \Delta\rho_{\rm CO_2}(t_1) \cdot \exp\left(\frac{t - t_1}{\tau_{\rm CO_2}} \right),
  \label{rhoCO2}
 \end{equation}
where: $\rho_{\rm CO_2}(t_0)\simeq 280$~ppm
 - pre-industrial concentration of CO\textsubscript{2};
 $\Delta\rho_{\rm CO_2}(t_1)\simeq 86.34$~ppm
 - empirically observed addition of CO\textsubscript{2}
by the year 2000 ($t_1=2000$~yr);
$\tau_{\rm CO_2}\simeq 46.8$~yr
 - time constant of CO\textsubscript{2}concentration growth in the post-industrial period. Using (\ref{rhoCO2}), the
expression (\ref{dotDelTH2Oq}) for $\frac{d}{dt} \Delta T(t)$
can be integrated over time for the growth of MAT due to the increase
in CO\textsubscript{2}. Indeed, if $\Delta T(t)$
is:
 \begin{equation}
 \Delta T(t) = \Delta T(t_1) \cdot \exp\left(\frac{t - t_1}{\tau_{\rm CO_2}} \right),
  \label{DelTCO2}
 \end{equation}
we can substitute it and also the expression for $\rho_{\rm CO_2}(t)$
 (\ref{rhoCO2}) into (\ref{dotDelTH2Oq}):
\begin{equation}
 \frac{d}{dt} \Delta T(t)
 =  \frac{\Delta T(t_1)}{\tau_{\rm CO_2}} \cdot \exp\left(\frac{t - t_1}{\tau_{\rm CO_2}} \right) \simeq
  \frac{1}{k\tau'} \left( k \cdot \sigma'_{\rm CO_2} \frac{\Delta\rho_{\rm CO_2}(t_1)}{\rho_{\rm CO_2}(t_0)}
   - \Delta T(t_1)  \right) \cdot \exp\left(\frac{t - t_1}{\tau_{\rm CO_2}}  \right),
 \label{dotDelTH2Or}
\end{equation}
  and this will establish the connection between $\Delta T(t_1)$
and $\rho_{\rm CO_2}(t_1)$:
\begin{equation}
\Delta T(t_1) \simeq \frac{\tau_{\rm CO_2}\sigma'_{\rm CO_2}}{\tau'+\tau_{\rm CO_2}/k}
 \cdot \ln \left( \frac{\rho_{\rm CO_2}(t_0) + \Delta\rho_{\rm CO_2}(t_1)}{\rho_{\rm CO_2}(t_0)} \right) .
 \label{DelTCO2t1}
\end{equation}
We note that the expression (\ref{DelTCO2t1}) is true for any values of $t_1$
 and can be considered as a universal relationship between $\Delta T(t_1)$
 and $\Delta\rho_{\rm CO_2}(t_1)$
 analogous to (\ref{DeltaT}). Comparing (\ref{DelTCO2t1}) and (\ref{DeltaT})
 allows us to write down the
expression (\ref{sigmaIPCC}) below, which gives us an estimate for $k$
 (Table~\ref{tab:table3}):
\begin{equation}
\sigma_{\rm IPCC} = \frac{\tau_{\rm CO_2}\sigma'_{\rm CO_2}}{\tau'+\tau_{\rm CO_2}/k}
\quad \Rightarrow \quad   k = \frac{\tau_{\rm CO_2} \sigma_{\rm IPCC} }
  {(\tau_{\rm CO_2} \sigma'_{\rm CO_2} -\tau' \sigma_{\rm IPCC} )} \simeq 10 .
 \label{sigmaIPCC}
\end{equation}

\begin{table}[b]
\caption{\label{tab:table3}%
Key parameters needed to estimate coefficient $k$
 and $\tilde{k}$.
}
\begin{ruledtabular}
\begin{tabular}{cccccc}
& $\tau_{\rm CO_2}$ & $ \tau' $ & $\sigma'_{\rm CO_2} $ & $ \sigma_{\rm IPCC} $ & $k$ \\
& (yr)                & (yr)     & (K)                 &  (K) &    \\
\colrule
max $k$ &  46.8  &  18.1  &  1.03 &  6.5  &  $ \infty$  \\
 $k$    &  46.8  &  9.44  &  1.31 &  4.33  &  9.9  \\
min $k$ &  46.8  &  5.78  &  1.58 &  2.16  &  1.64 \\
\colrule
max $\tilde{k}$ &  46.8  &  16.8  &  1.31 &  2.71  &  8     \\
 $\tilde{k}$    &  46.8  &  9.44  &  1.31 &  2.71  &  3.55  \\
min $\tilde{k}$ &  46.8  &  5.78  &  1.38 &  2.71  &  2.6   \\
\end{tabular}
\end{ruledtabular}
\end{table}

\subsection{Resulting renormalized values of the three key parameters
of the non-steady state model of global climate temperature change}

To sum up the previous section, the following values are estimated for
the three key parameters of the non-stationary model of global
temperature change (\ref{model}):
\begin{subequations}
\begin{equation}
 \sigma_{\rm CO_2} = k \cdot \sigma'_{\rm CO_2} \approx 13.1\;\mbox{K}
\end{equation}
\begin{equation}
 \sigma_a = k \cdot \sigma'_a \approx 635\;\mbox{K}
\end{equation}
\begin{equation}
 \tau = k \cdot \tau' \approx 100\;\mbox{yrs}
\end{equation}
\end{subequations}

To improve our estimates for $\sigma_{\rm IPCC}$, we can use the instrumental temperature
observations (Fig. 1C) for $\Delta T$(MAT).
Using regression, we estimate $\sigma_{\rm IPCC}=2.71$~K, which allows a more exact
value for $\tilde{k}$
 (Table~\ref{tab:table3}). We note that such more exact estimation of $\sigma_{\rm IPCC}$
can only be done within our approach, because expression (\ref{DeltaT}) is only a
particular case of the more general non-stationary model of global
temperature change (\ref{model}) within the period of rapid, quasi-exponential
growth of both GHG concentrations and MAT (red curves on Fig. 1 B and
C). In the traditional stationary approach the same model (\ref{DeltaT}) required
different estimates for $\sigma_{\rm IPCC}$.

The obtained more exact values of $\tilde{k}$
 allow to calculate a few scenarios of the future global temperature
change based on the non-stationary model (\ref{model}), for example under radical
reduction of anthropogenic GHG emissions by half in 2050 and 90\% by
2100. Fig. 1A shows the rate of increase of CO\textsubscript{2} under such
scenario. This would allow to stabilize CO\textsubscript{2} at about 500
ppm (Fig. 1B, grey curve), if assuming constant albedo and solar
forcing $\Delta S(t)\equiv \Delta a(t) \equiv 0$, the main equation (\ref{model}) is:
\begin{equation}
\frac{d}{dt}  \Delta T(t) = \frac{1}{\tau} \cdot \left( \sigma_{\rm CO_2} \cdot
\ln\left( \frac{\rho_{\rm CO_2}(t)}{\rho_{\rm CO_2}(t_0)}  \right)
- \Delta T(t)  \right) .
\label{modelB}
\end{equation}
Depending on the estimate of $\tilde{k}$
 (Table~\ref{tab:table3}), using formulae 22a-c we can obtain different temperature
growth estimates (Fig. 1C, grey curves 1, 2, and 3).

\begin{figure}[b]
\includegraphics[width=0.7\linewidth]{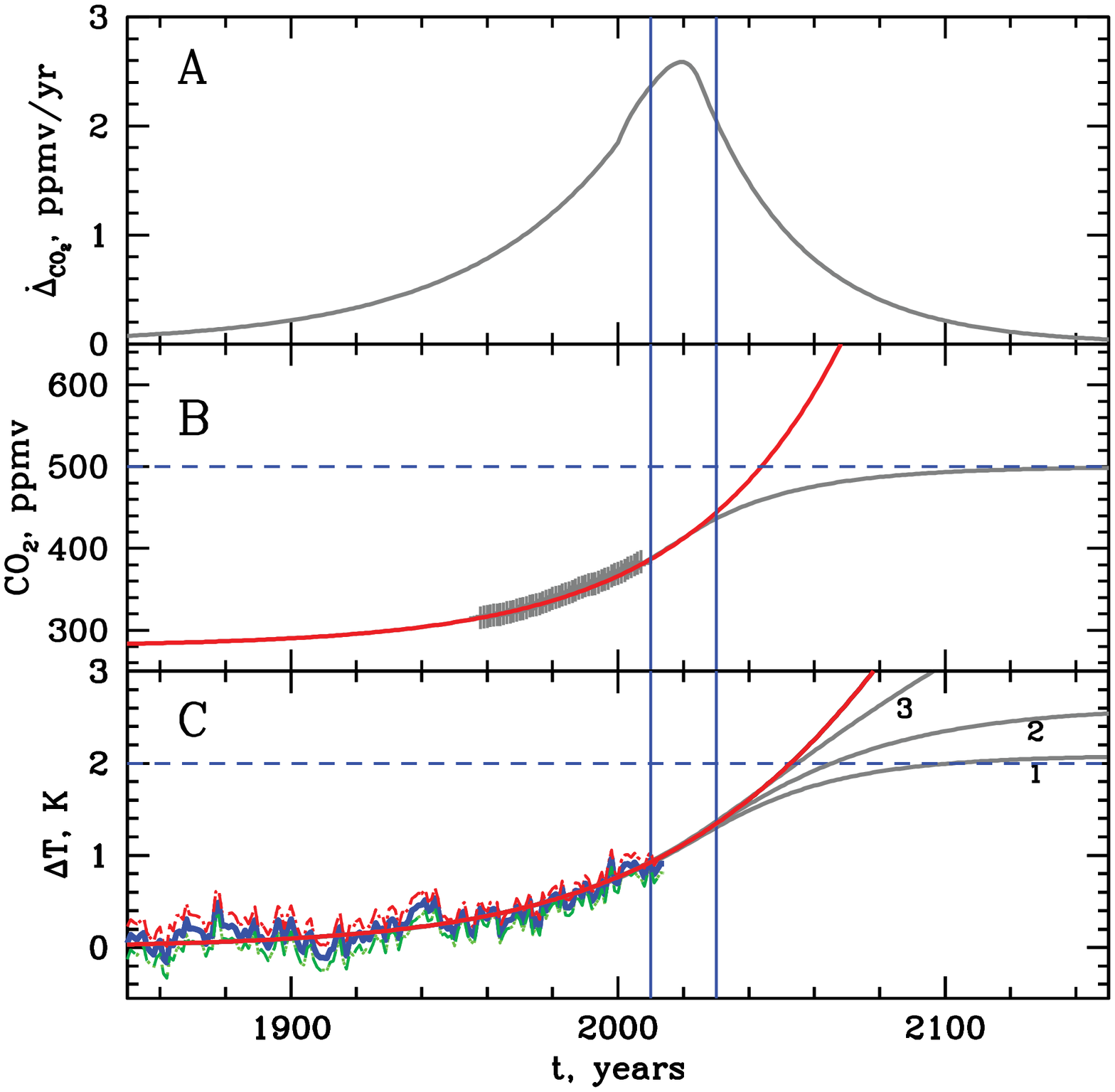}
\caption{\label{alltogether}
A. Annual increase of CO\textsubscript{2}
concentration under drastic anthropogenic GHG reduction scenario (50\%
by 2050 and 90\% by 2100 relative to 2010).
\newline
B. CO\textsubscript{2}
concentration change under uncontrolled (red curve) or limited
emissions (grey curve). Grey dots are MAT instrumental records from
NOAA ESRL.
\newline
C. Instrumental values of MAT \cite{Brohan2006,Met2016}
and future
temperature curves under various modeled scenarios using our model
(Model (\ref{modelB}) parameters:
$
 \sigma_{\rm CO_2} = \tilde{k}  \cdot \sigma'_{\rm CO_2} ; \;
 \tau = \tilde{k} \cdot \tau' ; \; \sigma'_{\rm CO_2} ; \; \tau' ; \; \tilde{k}
$
 from Table~\ref{tab:table3}).
}
\end{figure}

The main conclusion from our model is that the growth of global MAT is
most likely going to increase 2~K, and under some scenarios, rather
substantially (Curve 3 on Fig. 1C). This growth will continue due to
inertia in the model even under flat or reducing emissions of
CO\textsubscript{2} in the future due to non-linearity of the
temperature response in our model.

\section{\label{sec:Discussion} Discussion}
\subsection{\label{sec:MAT2K}Why MAT{\textgreater}2~K can be significant?}

An increase in MAT {\textgreater}
2~K  by 2100
is an alarming prospect \cite{IPCCWG1PhysicalStocker2013}. The danger is not only
the resulting increases in sea levels or biodiversity crisis. At stake
is the very survival of humanity, especially considering the extreme
rapidity of change. Non-linear feedbacks in climate systems and
probable existence of yet unknown critical thresholds
\cite{Karnaukhov1994,Alley2003,Schneider2004} makes any direct predictions problematic. Our
model highlights the importance of constraining the available fluxes of
carbon in limestone carbonates (over $3 \cdot 10^{7}$~Gt), ocean
pool CO\textsubscript{2} ($4 \cdot 10^{4}$~Gt), methane
hydrates ($0.5 \mbox{ to } 2.5 \cdot 10^{3}$~Gt), permafrost organics
({\textless}0.5~Gt), peat ({\textless}0.5~Gt), soil organics
({\textless}0.5~Gt) and wood biomass (about 0.6~Gt). One of the
understudied feedbacks is the influence of temperature rise on trees
worldwide, both temperate and tropical and, in turn, the shifting
albedo values on climate as a result of tree species replacement
\cite{Hollinger2010}.
For boreal forests of North
America and Eurasia one mitigation measure may be preventative planting
of fast growth species of broadleaf forest plantations (e.g., GMO
poplar or aspen) in lieu of conifers \cite{Larocque2013}.

\subsection{Fast growth broad-leaf tree plantations and their possible
role in climate mitigation}

Such plantations were initially proposed as a way of dealing with
increasing fire threats in the warming climate. However, they can also
be an efficient way of rapid carbon sequestration. For example, if
their total area reaches $A_f= 5$ million km\textsuperscript{2},
which is about 1\% of the Earth land
surface, a realistic value for the boreal forest zone of the Earth, we
may assess the total amount of sequestered carbon annually using (\ref{Mcarbon})
below.

Let us denote by $P$ the production of lumber by these plantations. It was shown that $P$ is
approximately 20 m\textsuperscript{3} ha\textsuperscript{{}-1} year\textsuperscript{{}-1} or $2 \cdot 10^3 $ m\textsuperscript{3} km\textsuperscript{{}-2} year\textsuperscript{{}-1} ($P$) 
(1 km\textsuperscript{2}=100 ha) \cite{Shestibratov2011,Vidyagina2014,Shanin2015}.

Considering that one m$^3$ of
wood contains approximately 200~kg of carbon $\chi_C \approx 200 kg/m^3$, about 2 Gt/year of carbon could be thus sequestered we denote this quantity as($M_C$):

\begin{equation}
M_C = \chi P S_f \simeq 200 \, \frac{\mbox{kg}}{\mbox{m}^3} \cdot 2\cdot 10^3
\frac{\mbox{m}^3}{\mbox{km}^2 \cdot \mbox{yr}} \cdot 5 \cdot 10^6 \, \mbox{km}^2 =
2 \cdot 10^{12}\, \frac{\mbox{kg}}{\mbox{yr}} = 2 \, \mbox{Gt/yr} .
 \label{Mcarbon}
\end{equation}

Two Gt of carbon equals about 20\% of modern anthropogenic carbon
emissions. Another benefit is the change in albedo of the boreal zone,
with the estimates of broad-leaf native albedo values averaging 0.15,
while coniferous forests average about 0.1 \cite{Leonardi2015}
(a difference of almost 35\%). Matthies and Valsta \cite{Matthies2016}
provided estimates of the difference between coniferous and
deciduous temperate forest albedo ranging from 0.02 to 0.07.

Thus, by
replacing coniferous forests with deciduous leaf tree plantations the
albedo of the surface increases by about 50\%. Considering that the
spherical albedo of the Earth is approximately $a=0.3$,
and the terrestrial land surface contributes 25\% of all outgoing
short-wave radiation, we can roughly estimate the change in albedo due
to plantations as

\begin{equation}
 \Delta a \simeq a \cdot 25\% \cdot 1\% \cdot 50 \% = 3.75 \cdot 10^{-4} .
 \label{DeltaA}
\end{equation}

Applying our parameters from Table III to the non-stationary model (28) we
may estimate the impact of albedo change on the asymptotic value of MAT
as

\begin{equation}
\Delta T = \tilde{k} \sigma'_a \left(\frac{\Delta a(t)}{1-a} \right) \simeq
   2.6\,(3.55,\; 8) \cdot 63.5\,\mbox{K}\cdot\left(\frac{3.75\cdot 10^{-4}}{1-0.3}\right)
   = 0.09\,(0.12,\;0.27)\,\mbox{K} .
 \label{DelTalb}
\end{equation}

The main conclusion from this consideration is that the change in albedo due to
fast-growing deciduous tree plantations may trigger a
constraining effect on the global temperature rising from
10\% ($\tilde{k}=2.6$) to 30\% ($\tilde{k}=8$) less as compared
to the current rate of temperature increase with respect to the
pre-industrial level. These would-be fast-growing tree plantations could potentially
mitigate climate change by providing bio-fuels, reducing
fire risks, and decreasing overall Earth surface albedo in
the vast boreal zone.

\section{Summary}
In this paper we propose a non-steady state model of global temperature change as
a baseline for a zero-dimensional climate model. Main equations of the model
were derived considering an impact of water vapor feedback. The
albedo dependent climate sensitivity and the relaxation time constant of
Earth's climate system were introduced and estimated on the basis of the
derived equations.

It is highly likely that our approach may improve existing 3D General Circulation
models. Our model asserts ongoing increase of the mean average temperature even after
the concentration of(CO\textsubscript{2} presumably stopped or even began to decrease. This increases
the probability of the global temperature exceeding 2 K limit even under aggressive
constraining of greenhouse gases.
Projects, such as planting rapidly growing deciduous trees in the boreal zone of Northern
Hemisphere may prove beneficial to offset such temperature rising because of the albedo
feed-backs. We provide an estimate of temperature increase slowdown because of the potential
albedo variations.
Our model is conceptually simple and thus could provide solutions for Climate
community, international organizations, governments and beyond.

\section{Acknowledgments and Data}
\begin{itemize}
\item
This work is supported by the Ministry of Education and Science of the
Russian Federation (Project no. 14.616.21.0013 from 17.09.2014, unique
identifier RFMEFI61614X0013).
\item
There are no real or perceived financial conflicts of interests for any
author of this article.
\item
We used publicly available datasets from the IPCC 5\textsuperscript{th}
Assessment
\url{http://ipcc.ch/publications_and_data/publications_and_data_other.shtml}
and NASA \url{http://data.giss.nasa.gov/}
\end{itemize}

\bibliography{karn}

\end{document}